\documentclass[12pt,a4paper]{article}
\pdfoutput=1
\usepackage{jheppub}
\usepackage{amsfonts}
\usepackage{bbm}
\usepackage{graphicx}
\usepackage[utf8]{inputenc}
\usepackage[T1]{fontenc}
\usepackage{amssymb,amsfonts,amsmath, mathtools}

\font\cmss=cmss10
\font\cmsss=cmss10 at 7pt
\font\manual=manfnt

\newcommand{\bi}{\begin{itemize}}
\newcommand{\ei}{\end{itemize}}
\newcommand{\non}{\nonumber}
\newcommand{\bea}{\begin{eqnarray}}
\newcommand{\eea}{\end{eqnarray}}
\newcommand{\be}{\begin{equation}}
\newcommand{\ee}{\end{equation}}
\newcommand{\ben}{\begin{eqnarray*}}
\newcommand{\een}{\end{eqnarray*}}
\newcommand{\bem}{\begin{pmatrix}}
\newcommand{\eem}{\end{pmatrix}}
\newcommand{\bl}{\begin{align}}
\newcommand{\el}{\end{align}}
\newcommand{\beg}{\begin{gather}}
\newcommand{\eeg}{\end{gather}}

\newcommand{\cA}{\mathcal{A}}
\newcommand{\cB}{\mathcal{B}}
\newcommand{\cC}{\mathcal{C}}
\newcommand{\cD}{\mathcal{D}}

\newcommand{\cI}{\mathcal{I}}

\newcommand{\cO}{\mathcal{O}}
\newcommand{\cS}{\mathcal{S}}

\newcommand{\cR}{\mathcal{R}}

\newcommand{\IH}{\mathbb{H}}

\renewcommand{\a}{\alpha}
\renewcommand{\b}{\beta}
\renewcommand{\d}{\delta}
\newcommand{\e}{\epsilon}

\renewcommand{\l}{\lambda}
\newcommand{\m}{\mu}
\newcommand{\n}{\nu}

\renewcommand{\r}{\rho}                                     
\newcommand{\s}{\sigma}

\newcommand{\D}{\Delta}

\renewcommand{\O}{\Omega}

\newcommand{\half}{\frac{1}{2}}
\newcommand{\pa}{\partial}
\newcommand{\nn}{\nonumber}
\newcommand{\Tr}{\mbox{Tr}}
\newcommand{\tr}{\mbox{tr}}

\def\dbend{\lower3.5pt\hbox{\manual\char127}}

\def\IL{\relax{\rm I\kern-.18em L}}
\def\IH{\relax{\rm I\kern-.18em H}}
\def\rlx{\relax\leavevmode}

\def\ZZ{\rlx\leavevmode\ifmmode\mathchoice{\hbox{\cmss Z\kern-.4em Z}}
 {\hbox{\cmss Z\kern-.4em Z}}{\lower.9pt\hbox{\cmsss Z\kern-.36em Z}}
 {\lower1.2pt\hbox{\cmsss Z\kern-.36em Z}}\else{\cmss Z\kern-.4em
 Z}\fi}

\title{Nonlocal Quantum Effective Actions  in Weyl-Flat Spacetimes}

\author[1]{Teresa Bautista,}
\author[2, 3]{Andr\'e Benevides,}
\author[3, 4, 5]{Atish Dabholkar}

\affiliation[1]{Max Planck Institute for Gravitational Physics (Albert Einstein Institute)\\
M\"{u}hlenberg 1, D-14476 Potsdam, Germany}

\affiliation[2]{SISSA, Via Bonomea 265, Trieste 34136 Italy}

\affiliation[3]{International Centre for Theoretical Physics\\
Strada Costiera 11, Trieste 34151 Italy}

\affiliation[4]{Sorbonne Universit\'es, UPMC Univ Paris 06\\
  UMR 7589, LPTHE, Paris, F-75005 France}
  
\affiliation[5]{CNRS, UMR 7589, LPTHE,  Paris, F-75005 France}

\abstract{Virtual massless particles in quantum  loops lead to nonlocal effects  which can have  interesting consequences, for example, for  primordial magnetogenesis in cosmology or for computing finite $N$ corrections in holography. We describe how  the quantum effective actions summarizing these effects can be  computed efficiently for Weyl-flat metrics by integrating the Weyl anomaly or, equivalently, the local renormalization group equation. This  method   relies only on the local Schwinger-DeWitt expansion of the heat kernel and allows for  a re-summation of leading large logarithms in situations where the Weyl factor changes by several e-foldings.  As an illustration, we obtain the quantum effective action for the Yang-Mills field coupled to massless matter, and the self-interacting massless scalar field. Our action reduces to the nonlocal  action obtained using the Barvinsky-Vilkovisky covariant perturbation theory in the regime $R^{2} \ll \nabla^{2} R  $ for a typical curvature scale $R$, but  has a greater range of validity effectively re-summing the covariant perturbation theory to all orders in curvatures. In particular, it is applicable also in  the opposite regime $R^{2} \gg \nabla^{2} R$, which is  often of interest in cosmology.}

\begin{document}
\maketitle

\section{Introduction}

Virtual massless particles in quantum  loops lead to nonlocal effects.
The quantum  dynamics of such massless particles coupled to a slowly evolving metric is summarized by the  one-particle-irreducible (1PI) quantum effective action for the background fields obtained by  integrating out the quantum  loops. Unlike the Wilsonian effective action, the 1PI effective action necessarily contains nonlocal terms which are not derivatively suppressed. These nonlocal terms can have  interesting consequences,  for example, for  primordial magnetogenesis in cosmology or for computing finite $N$ corrections in $AdS/CFT$  holography.

The computation of  the nonlocal quantum effective action is in principle a well-posed problem in perturbation theory. One can regularize the path integral covariantly using dimensional regularization or  short proper-time regularization  and evaluate the effective action using the background field method. However, explicit evaluation of  the path integral is forbiddingly difficult. For instance, to obtain the one-loop effective action it is necessary to compute the heat kernel of a Laplace-like operator in an arbitrary background, which amounts to solving the Schrödinger problem for an arbitrary potential. For short proper time, the heat kernel can be computed using the  Schwinger-DeWitt expansion \cite{Schwinger:1951xk, DeWitt:102655} which is analogous to the high temperature  expansion. This  is adequate for renormalizing the local ultraviolet divergences and to obtain the Wilsonian effective action if the proper time is short compared to the typical radius of curvature or the Compton wavelength of the particle being integrated out. However,  the nonlocal 1PI effective action receives contributions from the entire range of the proper time integral and the Schwinger-DeWitt expansion is in general not adequate. 
 
To obtain the full nonlocal effective action, one could  use the covariant nonlocal expansion of the heat kernel developed by Barvinsky, Vilkovisky, and collaborators \cite{Barvinsky:1984jd,Barvinsky:1985an}.  The effective action in this expansion has been worked out  to third order in curvatures in a series of important papers \cite{Barvinsky:1988ds,Barvinsky:1994hw,Barvinsky:1994cg,Barvinsky:1995it} and illuminates a number of subtle issues, for example, concerning  anomalies and the Riegert action \cite{Riegert:1984kt,Deser:1996na,Erdmenger:1996yc,Erdmenger:1997gy,Deser:2000un}. However, for a general metric the explicit expressions   are rather complicated already at the third order.  Furthermore,  the Barvinsky-Vilkovisky (BV) expansion requires $\cR^{2} \ll \nabla^{2} \cR  $, where $\cR$ denotes a \textit{generalized} curvature including both a typical geometric curvature $R$  as well as a typical gauge field strength $F$. One is often interested though in the regime of slowly varying curvatures,  $R^{2} \gg \nabla^{2} R$, for example during slow-roll inflation. This is beyond the validity of the BV regime. 

The aim of the present work is to find  practical methods to go beyond these limitations but only for a restricted class of metrics that are Weyl-flat and for classically Weyl invariant actions. In this case, one can exploit the symmetries of the problem.  The only dynamical mode of the background metric is the Weyl factor which is a single function. The Weyl anomaly is the Weyl variation of the action which  can be viewed as  a first order  scalar functional equation for the action that can be easily integrated.  The initial value of the  action functional  can often be  determined by the flat space results. In this manner, the entire effective action including its anomalous dependence on  the Weyl factor  can be determined efficiently. 

The main advantage of our approach is that one can extract  the essential physics with relative ease.  Weyl anomalous dimensions of local operators (or equivalently the beta functions) can be  computed reliably using  \textit{local} computations such as the Schwinger-DeWitt expansion.  The  resulting actions are necessarily nonlocal  much like the Wess-Zumino action for chiral anomalies\footnote{The chiral anomaly itself can be deduced from local Schwinger-DeWitt expansion. The nonlocal Wess-Zumino action is  then obtained by the Wess-Zumino construction which essentially integrates the local anomaly equation. Our method  extends this procedure to situations with nontrivial beta functions.}.  Even though   we relax the restriction $R^{2} \ll \nabla^{2} R $, we still  need to assume $F^{2} \ll \nabla^{2} F  $ for a typical field strength $F$. 
In summary, the Barvinsky-Vilkovisky regime requires \textit{rapidly varying curvature} as well as \textit{rapidly varying field-strength} whereas our regime requires only rapidly varying field-strength.
Our method essentially re-sums the BV  expansion to all orders in curvatures albeit for a restricted class of Weyl-flat metrics as we discuss in \S\ref{BV}. 
 
These nonlocal actions for Weyl-flat metrics can have a number of interesting applications. In $AdS/CFT$ correspondence, Weyl-flat metrics are relevant for the bulk description of  renormalization group flows in  the boundary CFT. Loop effects of massless supergravity fields are important, for example,  in the computation of finite $N$ effects in the bulk such as the finite charge corrections to the Bekenstein-Hawking entropy of black holes \cite{Dabholkar:2011ec, Dabholkar:2014ema,Sen:2011ba,Banerjee:2011jp}.  In cosmology, the Robertson-Walker metric  for an isotropic and homogeneous universe with flat spatial section is Weyl flat. 
During many epochs in the early universe, various particles can be massless or nearly massless compared to the Hubble scale. Quantum loops of these particles can lead to an  anomalous dependence on the Weyl factor which can have interesting consequences. For example, in massless electrodynamics it can contribute to the generation of primordial magnetic fields \cite{Turner:1987bw,Ratra:1991bn,Demozzi:2009fu,El-Menoufi:2015ztk,Benevides:2017b} where one is precisely in the regime
of rapidly varying field strengths but slowly varying curvatures. 
This approach  can also be useful for exploring the stability of de Sitter spacetime, and the cosmological evolution of the Weyl factor and other physical  parameters in  quasi de Sitter spacetimes in four dimensions similar to the  two-dimensional models analyzed in \cite{Dabholkar:2015qhk, Bautista:2015wqy, Bautista:2015nxc}. Possible implications of nonlocal actions have been explored, for example,  in 
\cite{,Mottola:2006ew,Deser:2007jk,Park:2012cp,Woodard:2014iga,Donoghue:2014yha,Godazgar:2016swl}. 

Our method naturally lends itself to a re-summation of leading large logarithms  using a `local renormalization group improvement' as we discuss in \S\ref{Scalar}. In cosmology, one is dealing with extremely long times and distances spanning several logarithmic scales. The scale factor of the universe underwent at least  $60$ and possibly more $e$-foldings. Thus, large logarithms could render perturbation theory invalid. These logarithmic quantum corrections  add up to something appreciable and have a potential for interesting applications if they can be properly re-summed. 

The paper is organized as follows.
In  \S\ref{Weyl}, we describe how the Weyl anomaly can be used  to determine the quantum effective action with the help of an `integration lemma’ and the Schwinger-DeWitt expansion of the heat kernel. In \S\ref{NEA} we apply this method to compute the effective action for  the Yang-Mills field coupled to massless matter and for a  self-interacting massless scalar field.   In \S\ref{BV}, we discuss the relation of our results to  the nonlocal covariant perturbation theory developed by Barvinsky-Vilkovisky  and similar results obtained by Donoghue and El-Menoufi \cite{Donoghue:2015xla,Donoghue:2015nba} using Feynman diagrams in the weak field limit. 

\section{Effective Actions from Weyl Anomalies \label{Weyl}}

In this section we describe the general method for computing the quantum effective action at the one-loop order for essentially all the standard model fields in Weyl-flat spacetimes by integrating the Weyl anomaly. To simplify the discussion, we ignore Yukawa couplings and work  in the conformal massless limit so that all couplings are dimensionless. Dimensionful couplings and non-conformal scalars can be incorporated  with some modifications \cite{Benevides:2017a}.   We first review  elements of the background field method and gauge fixing  to set up our conventions. We then discuss the anomalies in terms of the Schwinger-DeWitt expansion and a lemma to obtain the effective action by integrating the anomaly.

\subsection{Classical actions and the Background Field method}

 Consider the classical  action for a conformally coupled real scalar field $\varphi$ with quartic self-interaction:
\be\label{c-action-scalar}
\cI_{0}[g, \varphi] = - \int d^{4}x \sqrt{|g|}  \left[  \frac{1 }{2} |\nabla \varphi |^{2}+ \frac{1}{12} R\, \varphi^{2}+ \frac{\l_{0}}{4!} \varphi^{4}   \right] \, ,
\ee 
where $\l_{0}$ is the bare coupling and $R$ the Ricci scalar for the metric $g$. This can also be viewed  as the bare action in the ultraviolet if we regard the fields as bare fields. 
Even though we are interested in the Lorentzian action, for subsequent computations it is convenient to use the Wick-rotated action  on the Euclidean section:
\be\label{c-action-scalar2}
\cS_{0}[g, \varphi] =  \int d^{4}x \sqrt{|g|}  \left[  \frac{1 }{2} |\nabla \varphi |^{2}+ \frac{1}{12} R\, \varphi^{2}+ \frac{\l_{0}}{4!} \varphi^{4}   \right] \, .
\ee 

We denote the Lorentzian action by $\cI$ and the Euclidean action by $\cS$. Wick rotation of Lorentzian time $t$ to Euclidean time $t_{E}$ can be thought of as a coordinate change  $t= -it_{E}$ in the complexified spacetime.  Tensors transform as tensors under this coordinate change and in particular the Lagrangian transforms as a scalar. The path integral is defined with  weight $e^{i\cI_{0}}$ in Lorentzian spacetime but  with $e^{-\cS_{0}}$ in Euclidean space.  Using the fact that the volume element $\sqrt{|g|}$ equals $\sqrt{-g}$ on Lorentzian section but $\sqrt{g}$ on the Euclidean section, the two actions are simply related by $\cI[g, \varphi] \rightarrow -\cS [g, \varphi]$ as above.

In the background field method  \cite{Abbott:1981ke}, one splits the quantum field as $\hat\varphi = \varphi + Q$, a sum of a background field $\varphi$ and the quantum fluctuations $Q$ around this background.
The quantum effective action $\cS[\varphi]$ for the  background field $\varphi$ is then given by the path integral
\be
\exp{\left(-\cS[g, \varphi]\right)} := \int \cD Q \, \exp { \left(
-\cS_{0}[g, Q + \varphi] - J[\varphi]  Q   \right) } \, ,
\ee
where the external current 
\be
J[\varphi] (x)= \frac{\d \cS[\varphi]}{\d \varphi(x)} = \frac{\d \cS_{0}[\varphi]}{\d \varphi(x)} + \ldots 
\ee
is a function of the background field  adjusted so  that the tadpoles vanish order by order in perturbation theory. We use  short-proper time cutoff as a manifestly covariant regulator in the heat kernel method as described below. In renormalized perturbation theory, the UV divergences are renormalized with appropriately chosen  counter-terms and all physical quantities are expressed in terms of the renormalized coupling $\l$ defined at a mass scale $M$. At one-loop order, the path integral can be approximated  by the Gaussian functional integral 
\be
e^{-\cS_{1}[g, \varphi]}=  e^{-\cS_0[g, \varphi]}\int \cD Q  e^{-\half  \, \langle Q | \cO_\varphi |Q\rangle} \, ,
\ee
where  $\cO_{\varphi}$ is the quadratic fluctuation operator in the background:
\be
\cO_{\varphi} = -\nabla^2 + \frac{\l \varphi^{2}}{2} + \frac{1}{6} R \, 
\ee
with
\be
 \nabla^{2} =  \frac{1}{\sqrt{| g |}}\partial_{\m} (\sqrt{|g|}\, g^{\m\n} \partial_{\n}) \,  .
\ee
The Gaussian integral can be evaluated in terms of the determinant of $\cO_{\varphi}$,  
\be
\cS_{1} = \cS_0  + \half \log \det \left(\cO_\varphi\right) = \cS_0  +  \half  \Tr \log\left(\cO_{\varphi}\right) \, .
\ee
We use the convention
\be\label{Tr}
\int \,   d^{4}x \sqrt{|g|}\, |x\rangle\langle x| \,= \mathbf{1}\, ; \qquad 
\Tr ( \cO)  = \int d^{4}x \sqrt{|g|}\, \langle x| \,\tr\, \cO  \,|x\rangle \, .
\ee

We  next consider gauge theory, concretely an $SU(N)$  Yang-Mills field coupled to a massless complex scalar and a  massless Dirac fermion transforming in the fundamental representation. The  classical Lorentzian action is
\be\label{action-L}
\cI_0[g, A] =  - \int   d^4x \, \sqrt{|g|} \, \left[\frac{1}{4\,e_0^2}  \, F^{a}_{\mu\nu} F^{a\mu\nu} 
+ \left|D\Phi\right|^2  +\frac{1}{6}R\left|\Phi\right|^2 +  i \,\bar \Psi\, \Gamma^\a\, e_\a^\mu \,  D_\mu\Psi \right] \, ,
\ee
where $e_0^{2}$ is the bare gauge coupling and   $a$  is the adjoint index $( a= 1, 2, \ldots ,N^{2}-1 )$. 
The covariant derivative is now defined including both the spin and the gauge connection:
\be
D_{\mu}:=  \partial_{\mu} + \half w_{\mu} ^{\a\b} J_{\a\b}+ A^{a}_{\mu} \,T_{a}\, , \qquad  (\a, \b = 0, \ldots,  3)\,  ,
\ee
where $\{J_{\a\b}\}$ are the Lorentz representation matrices and
$\{ T_{a} \}$ are the anti-Hermitian $SU(N)$ representation matrices   normalized so that $\tr_{F} ({T_{a} T_{b}}) = -\half \delta_{ab}$ in the fundamental representation $F$.
The quantum field $\hat{A}_\mu$ is a sum of a  background $A_\mu$ and a  quantum fluctuation $a_\mu$, 
$\hat{A_{\m}} = A_{\m} + a_{\m}$.
To choose the background gauge, the gauge transformation of the quantum gauge field 
\be
\delta_{\e} \hat{A}_\mu := \hat D_{\mu} \e = \pa_\mu \e + [\hat{A}_\mu,\e]\, 
\ee
can be split as
\be
\delta_{\e} A_\mu = \pa_\mu \e + \left[A_\mu,\e\right] := D_{\m}\e \, ,\qquad\qquad
\delta_{\e} a_\mu = \left[a_\mu,\e\right]\, .
\ee
It is convenient to choose the background gauge $D_{\mu}a^{\mu}=0$ so that the effective action for the background field is manifestly gauge invariant. We set the background fields for $\Phi$ and $\Psi$ to zero. Following the standard Fadeev-Popov procedure we add the gauge fixing term and ghost Euclidean actions which at one-loop are of  the form
\be
\cS_{gf} = \frac{1}{2\,e_0^2\,\xi} \int   d^4x \, \sqrt{|g|} \, \left| D_\m \,a^\m\right|^2\, , \qquad\qquad \cS_{gh} = - \int   d^4x \, \sqrt{|g|} \, \bar{c}\,D^2 c \, 
\ee
where the covariant derivatives contain only the background connection. We henceforth use the ’t Hooft-Feynman gauge $\xi=1$.

The one-loop quantum effective action is then given by
\be\label{trlog1}
\cS_{1} = \cS_0 +  \Tr \log\left(\cO_\Phi\right)- \half\,\Tr \log\left(\cO_\psi\right) + \frac{1}{2}\,\Tr \log\left(\cO_{A}\right)- \Tr \log\left(\cO_{c}\right)\, .
\ee
The operators involved are typically of the second-order Laplace-type
 \be
\cO_{f}=-g^{\m\n}\,D_\m\,D_\n\, \textbf{1}+\textbf{E}\, ,
\ee
where $D_\m$ is the covariant derivative defined above which depends on the representation of the field, $\textbf{1}$ is the identity in the  representation space of the field, and $\textbf{E}$ is the `endomorphism matrix’ that depends on the background fields.

The regularized functional trace for various operators $\cO_{f}$ can be expressed in terms of the diagonal elements of the corresponding heat kernels 
$K_{f}(s) := e^{-s \,\cO_{f}}$
by the standard expression:
\bea\label{trlog2}
\Tr \log\left(\cO_f\right) = -\int\limits_\e^\infty \frac{ds}{s} \,\, \Tr \,K_{f}(s)  &=&  - \int\limits_\e^\infty \frac{ds}{s}\int d^4x \sqrt{|g|} \langle x| \tr \, K_{f}(s)|x \rangle \nonumber \\
&=& -\int\limits_\e^\infty \frac{ds}{s} \,\,\int d^4x \sqrt{|g|}\,  \tr \, K_{f}(x,x;s)\, .
\eea
Here `$\Tr $’ is a total trace  including the spacetime `index’ $x$ as in \eqref{Tr}  as well as the matrix indices of the Lorentz and $SU(N)$ representations,  whereas `$\tr$’ is a trace over only  the matrix indices\footnote{See for example \cite{Percacci:2017fkn, Vassilevich:2003xt} for notational conventions.}.
The short proper time cut-off $\e$ has mass dimension $-2$ and hence we can write $\e= M_{0}^{-2}$ and regard $M_{0}$ as the UV mass cutoff.

In general,  it is not possible to evaluate  $K_{f}(x,x;s)$ explicitly for all values of the proper time. 
However, exploiting  Weyl anomalies and the symmetries of Weyl-flat backgrounds, it is possible to compute $\cS$ avoiding the proper time integral altogether, as we discuss in the next two sections.

\subsection{Weyl Anomaly and the Local Renormalization Group}

Since regularization with a short proper time cutoff $\e$ is manifestly covariant, we do not expect any anomalies in the diffeomorphism invariance. On the other hand, the cutoff scale $M_{0}$  introduces a mass scale  and there is a potential for Weyl anomalies.

The  local Weyl transformation of the spacetime metric $g_{\m\n}$ is defined by
\be
\quad g_{\m\n} \rightarrow e^{2 \xi (x)} g_{\m\n} \, , \quad \,\, g^{\m\n} 
\rightarrow e^{-2 \xi (x)} g^{\m\n} \, , 
\ee
or infinitesimally,
\be
g^{\m\n}(x) \rightarrow  g^{\m\n}(x) - {2\,\xi (x)}\, g^{\m\n}(x) \,  .
\ee
All other  fields we denote collectively as $\{\chi_{f}\}$ which  transform with Weyl weights $\{\Delta_{f}\}$
\be
\chi_{f}(x) \rightarrow e^{-\Delta_{f }\xi(x) }\chi_{f}(x)  \, .
\ee
In particular,  in four dimensions, a conformally coupled  scalar field has Weyl weight $1$, a fermion field has weight $3/2$, a gauge field has weight $0$ so that the kinetic terms are scale invariant. 
The local Weyl group $\mathcal{G}$ is an infinite dimensional  abelian group with generators $\{J_x$\} acting on the space of fields\footnote{Dimensionful couplings  could be treated as additional `spurion’ scalar fields with Weyl dimensions  equal to their classical mass dimensions so that the classical action is rendered Weyl invariant. This more general situation will be discussed in \cite{Benevides:2017a}. In this case, the background fields $\{\chi_{f}\}$ will include also the spurion fields.}:
\be
J_x :=  -2 \, g^{\mu\nu}(x)\frac{\delta\quad}{\d g^{\m\n}(x)} - \Delta_{f}\, \chi_{f}(x)\frac{\d\quad }{\d \chi_{f}(x)}  \, . 
\ee
Treating the coordinate $x$ of the local scaling parameter $\xi(x)$ as a continuous index,  we can write an element  of this group as
\be
e^{\,\xi \cdot J} 
\ee
with the `summation' convention
\be
\xi \cdot J :=  \sum_x  \xi_x J_x := \int d^4x \,  \xi (x) J_x \, .
\ee
A Weyl-flat metric can be written as
\be\label{weyl-flat}
g_{\m\n} = e^{2\Omega} \eta_{\m \n} = e^{\O \cdot  J} (\eta)
\ee
and is on the Weyl-orbit of the flat Minkowski metric $\eta_{\m\n}$.

Weyl invariance of the classical action implies that
\be
J_{x }\left( \cS_{0}[g, \chi_{f}]\right) = 0 \, .
\ee
The cutoff $\e$ required for defining the quantum path integral breaks Weyl invariance.  Consequently the 1PI quantum effective action $\cS$ for the background fields is no longer Weyl invariant. The quantum violation of classical Weyl invariance can be expressed as an anomaly equation:
\be\label{anomaly}
J_{x } \left( \cS[g, \chi_{f}]\right) := \left(-2 \,g^{\mu\nu}\frac{\delta\quad}{\d g^{\m\n}(x)} - \Delta_{f }\, \chi_{f} \frac{\d \quad}{\d \chi_{f}(x)}\right) \left(\cS [ g, \chi_{f}] \right)= - \mathcal{A}(x) \sqrt{|g|} \, ,
\ee
where $\mathcal{A}(x)$ is the Weyl anomaly scalar\footnote{In conformal field theory, Weyl anomaly is usually understood to mean only  the `conformal anomaly’ in  curved spacetime at the conformal fixed point, arising from the Weyl non-invariance of the measure. We denote this anomaly by $\mathcal{C}(x)$. More generally, interactions  perturb the theory away from the fixed point and the nontrivial beta functions  generate a renormalization group flow. In this case, the Weyl anomaly $\mathcal{A}(x)$  includes  the `beta function anomaly' $\cB(x)$  in addition to  the conformal anomaly and thus $\mathcal{A}(x) = \cB(x) + \cC(x)$. This notation should not be confused with the Type-$A$  and  Type-$B$ classification  of anomalies \cite{Deser:1993yx}.}. Since the violation of the Weyl symmetry is a result of the short-distance regulator, one expects on general grounds that the anomaly $\mathcal{A}$ must be \textit{local} even though the 1PI action is generically nonlocal. In particular, it must admit a local expansion in terms of the background fields. 
The locality of the anomaly is of crucial importance. At one-loop, one can prove it explicitly and obtain a formula for the anomaly in terms of the local Schwinger-DeWitt expansion.  

We illustrate the general argument for the conformally coupled scalar field $\varphi$. The
infinitesimal Weyl variation of the quadratic action for the quantum fluctuation vanishes:
\be
\delta_{\xi} \langle Q | \cO_{\varphi} |Q\rangle =  \d_{\xi} \int d^{4}x \,\sqrt{|g|}\, Q(x)  \,\cO_{\varphi}\, Q(x) =0 \, .
\ee
Using the  Weyl transformations of $Q$ and the background metric $g_{\m\n}$ we conclude that 
 \be
 \d_{\xi} \cO_{\varphi} = -2\, \delta \xi(x) \,\cO_{\varphi}
 \ee
 up to boundary terms. The quadratic fluctuation operator $\cO_{\varphi}$ is thus covariant  under Weyl transformations with weight 2.
It then follows that
 \bea
 \delta_{\xi} \cS_{1}[g, \varphi] &=&   -\half \int\limits_\e^\infty \frac{ds}{s} \,\, \,\,  \Tr \,   \delta  e^{-s\cO_{\varphi}} = \half  \int\limits_\e^\infty {ds} \,\, \,\,  \Tr \,   (\delta \cO_{\varphi })  e^{-s\cO_{\varphi}} \\
 &=&   - \int\limits_\e^\infty {ds} \,\,   \int d^{4}x \,\sqrt{|g|}\,\, \d\xi(x)\, \,\langle x| \, \tr \,\cO_{\varphi } \,  e^{-s\cO_{\varphi}}| x \rangle \\
 &=&   \int\limits_\e^\infty {ds} \,\, \,\, \int d^{4}x \,\sqrt{|g|}\,\,\d\xi(x) \,\,\frac{d}{ds} \,   \langle x| \, \tr   \,e^{-s\cO_{\varphi}}| x \rangle\, .
    \eea
 Performing the $s$ integral we obtain\footnote{If  the operator $\cO_{\varphi}$ has no zero modes there is no contribution from the upper limit of the integral.}
 \be
\frac{ \delta \cS_{1}[g, \varphi]}{\d \xi(x)} = J_{x}\left(  \cS_{1}[g, \varphi]\right) =
 -   \langle x|\, \tr \,  e^{-\e\,\cO_{\varphi}} |x \rangle \sqrt{|g|}\, 
 = - \tr \,K_{\varphi}(x, x;\e) \sqrt{|g|}\,  . 
 \ee
A similar reasoning can be used for fermions since the Dirac action is Weyl invariant in all dimensions. For gauge fields, there is an additional subtlety because the gauge fixed action  and the ghost action are not separately Weyl invariant. However, one obtains  an analogous expression for the combined system of gauge and ghost fields \cite{Codello:2012sn}.   Both for fermions and the gauge-ghosts system, the quadratic operators have Weyl weight two. The action of the Weyl generator on the field space is thus given by
 \be\label{anomaly1}
J_{x } \left( \cS_{1} [g, \chi_{f}]\right) := -\cA(x) \sqrt{|g|} = - 2\,\sum_{f} \,n_{f} \,\text{tr}\, K_{f}(x,x;\e) \sqrt{|g|}\,
\ee
where $n_{f}$ is the coefficient of  $\Tr \log (\cO_{f})$ in \eqref{trlog1} consistent with our convention in \eqref{trlog2}. 
Thus, $n_{\Phi}=1$, $n_{\Psi}= -\half$,  $n_{c}= -1$, $n_{A}=n_{\varphi}= \half $.

Equation \eqref{anomaly1} shows  the anomaly  is determined entirely by the short proper time behavior of the heat kernel. Since the proper time cutoff $\e$ effectively provides a covariant short-distance cutoff in spacetime, the resulting anomaly $\cA(x)$ is local as promised.  Therefore, it must admit an expansion in terms of local fields $V_{i}(x)$: for the beta function anomaly,
\be
\cB(x) =  \sum_{i}\beta_{i} V_{i}(x)  \, ;
\ee
the $\cC(x)$ anomaly is  purely gravitational and has a similar expansion in terms of the local functionals of the metric such as the Euler density \cite{Capper:1973mv, Deser:1976yx,Duff:1993wm}.

The Weyl anomaly equation is closely related to the local renormalization group  \cite{Drummond:1977dg,Osborn:1991gm,Jack:1990eb} and  the coefficients $\b_{i}$ can be simply related to the usual beta functions. We illustrate this connection for Yang-Mills theory. The  Weyl variation of the action with respect to the Weyl factor $\O$ of the metric  \eqref{weyl-flat} 
is given by \eqref{YMtrace} at one loop:
\be\label{YManomaly}
J_{x}( \cS_{1} [g, A]) = \frac{\delta \cS_{1} [g, A]}{\delta \Omega(x)} =  -\cB(x) \sqrt{|g|} =   \frac{b}{4}   \,  F^{2} (x)  \sqrt{|g|}\, ,
\ee
where $b$ is given by \eqref{qcdb} and we have ignored  the purely gravitational $\cC(x)$ anomaly. To relate it to
the local renormalization group, we note that a Weyl scaling of the metric increases length scales or decreases mass scales. Hence we can regard $M(x):= Me^{\Omega(x)}$ to be the position-dependent local renormalization scale\footnote{This is true as long as one is dealing with `primary' fields such as  $g_{\m\n}$ or $F_{\m\n}$ which transform covariantly under Weyl transformation. In general,  `secondary' fields such as $R_{\m\n}$ or $\nabla_{\m}\varphi$ are also relevant, which contain derivatives of the primary fields. In this case, the Weyl transformations contain terms with derivatives of the Weyl factor $\Omega$ and the equality \eqref{Mrelation} holds only up to these derivatives \cite{Drummond:1977dg,Shore:1986hk,Shore:1990wp,Osborn:1987au,Osborn:1989td, Osborn:1991gm,Benevides:2017a}.}  $M(x)$.  
Therefore, 
 \be\label{Mrelation}
M(x) \frac{\d\quad}{\d M(x)} =  \frac{\d\quad}{\d\Omega(x)}  \, .
 \ee
If the  scale $M(x)$ is position dependent, then it is natural to regard all renormalized couplings to be also position-dependent expectation values of  nondynamical `spurion’ fields. For example, regarding, $1/e^{2} = \l_{e}(x)$ as position dependent, and using \eqref{Mrelation} and \eqref{YManomaly} we conclude that
\be
 \left[M(x)\frac{\d}{\d M(x)} +\b_{e}\frac{\d }{\d \l_{e}(x)}\right]\cS_{1}=0 \, ,
\ee
with 
\be\label{RG}
\b_{e}:=M \frac{d \l_{e}}{dM}= M \frac{d e^{-2}}{dM} = -b \, .
\ee
For constant $M(x)$, functional derivatives are replaced by ordinary derivatives and one recovers the usual position-independent `global' homogeneous renormalization group equation. 

More generally, the local renormalization group equation is best thought of as a Weyl anomaly equation \eqref{anomaly} with a local expansion for the anomaly $\cA$. 

\subsection{Integration of the Weyl Anomaly \label{Integration}} 

Our goal is to deduce the nonlocal quantum effective action $\cS[g,\chi_f]$ by integrating the local  Weyl anomaly.  Towards this end, we consider the following trivial identity\footnote{We thank Adam Schwimmer for this formulation.}
\be
e^{\,\xi \cdot J}  = \textbf{1} + \int_{0}^{1} dt \,e^{\,t\, \xi \cdot J} \,\, \xi \cdot J \, .
\ee
We wish to compute $\cS [g, \chi_{f}]$ for 
$(g, \chi_{f})$ on the Weyl-orbit of $(\bar g, \bar \chi_{f})$ with Weyl factor $\Omega(x)$:
\be\label{orbit}
(g, \chi_{f}) = e^{\O \cdot J} (\bar g, \bar \chi_{f})\, .
\ee
 Using the identity above we obtain
\bea
\cS [g, \chi_{f}] \equiv e^{\O \cdot J} \left( \cS [\bar g, \bar \chi_{f}] \right)  &=&  \left( \textbf{1}  + \int_{0}^{1} dt\,e^{t\, \O \cdot J } \, \O \cdot J \right)\left( \cS [\bar g, \bar \chi_{f}]\right)\\
 &=&  \cS [\bar g, \bar \chi_{f}]  + \int_{0}^{1} dt \,e^{t\, \O \cdot J} \, \O \cdot J \left(\cS [\bar g, \bar \chi_{f}] \right)\\
 &=&  \cS [\bar g, \bar \chi_{f}]  - \int_{0}^{1} dt \,e^{t\, \O \cdot J} \left(\int d^4x \,\O(x)\sqrt{|\bar g|} \, \mathcal{A} [\bar g, \bar \chi_{f}] (x) \right)
\eea
where we have used \eqref{anomaly} in the last line. Using \eqref{orbit} we then conclude\footnote{The argument $g$ of  the action $\cS[g, \chi_{f}]$ functional here refers to the covariant tensor $g_{\m\n}$ and not $g^{\m\n}$.}
\be\label{lemmaE}
\cS [g, \chi_{f}] =  \cS [\bar g, \bar \chi_{f}]  + \,  \cS_{\cA}[\bar g, \O, \bar \chi_{f}] \, ,
 \ee
 where 
 \be\label{lemmaE2}
 \cS_{\cA}[\bar g, \O, \bar \chi_{f}] :=-\int_{0}^{1} dt \int d^4x \sqrt{|\bar g \,e^{2\, t\, \O(x)}|} \,\, \O (x) \,\, \mathcal{A} [\bar g \,e^{2\, t\, \O}, \bar \chi_{f} \,e^{- \D_{f}\, t\, \O}](x)
 \ee
 is the contribution to the action from the anomaly. 
Lorentzian continuation of \eqref{lemmaE} gives a similar equation 
 \be\label{lemmaL}
\cI [g, \chi_{f}] =  \cI [\bar g, \bar \chi_{f}]  + \,  \cI_{\cA}[\bar g, \Omega, \bar \chi_{f}] \, 
 \ee
 but with $\cI_{\cA}$ given by
 \be\label{lemmaL2}
 \cI_{\cA}[\bar g, \O, \bar \chi_{f}] :=\int_{0}^{1} dt \int d^4x \sqrt{|\bar g \,e^{2\, t\, \O(x)}|} \,\, \O (x) \,\, \mathcal{A} [\bar g \,e^{2\, t\, \O}, \bar \chi_{f} \,e^{- \D_{f}\, t\, \O}](x)
 \ee
 because the anomaly scalar does not change sign under Wick rotation.
 
 Equation \eqref{lemmaE} is  a simple identity that follows essentially from the group structure of Weyl transformations. It is thus applicable  to any order in perturbation theory if we can compute the Weyl anomaly to that order.   To compute the effective action to  the one-loop order,  one can use the expression \eqref{lemmaE2} with the Weyl anomaly given in terms of the heat kernel as in \eqref{anomaly1}. Since the short-time expansion of the heat kernel is determined by the local Schwinger-DeWitt expansion, we see that \eqref{lemmaE} enables us to determine the entire quantum effective action for Weyl-flat background metrics knowing only the local expansion.

Note that the left hand side of \eqref{lemmaL} depends only on the physical metric whereas the right hand side \textit{a priori} depends on the fiducial metric $\bar g$ and $\Omega$ separately. It must therefore be true that the action on the right hand side exhibits  `fiducial Weyl \textit{gauge} invariance’ 
\be\label{fiducial}
\bar g \rightarrow e^{2\zeta(x)} \bar g\, , \qquad \Omega \rightarrow \O -\zeta(x) \, ,
\ee
under which the fiducial metric $\bar g$ transforms but the physical metric $g$ is invariant. This gauge invariance reflects the fact that  splitting  $g$ into $\bar g$ and $\Omega$ is ambiguous, and  all splits related by a fiducial gauge transformation are  physically  equivalent. 
The fiducial gauge invariance of the right hand side  of \eqref{lemmaL} is necessary to show that it depends only on the physical metric. 
As we explain in $\S\ref{BV}$, it is often far from obvious how the answer obtained using our method can be expressed covariantly entirely in terms of the physical metric. However, the procedure guarantees that this must be the case.

\subsection{Schwinger-DeWitt Expansion of the Heat Kernel}

The trace of the heat kernel admits a short proper time expansion as
\be
\Tr \,K_f(\e)=\int\limits_\mathcal{M} d^dx\sqrt{|g|}\, \,\,\text{tr}\, K_f(x,x;\e)=\int\limits_\mathcal{M} d^dx\sqrt{|g|}\,\frac{1}{(4\,\pi\e)^{d/2}}\,\sum_{n=0}^\infty a_n(x)\, \e^{n}\, .
\ee
The $a_n(x)$ are the Gilkey-Seeley-HaMiDeW \cite{Gilkey:1995mj,Seeley:1967ea,Seeley:1969re,Hadamard:2014le,Minakshisundaram:1949xg,Minakshisundaram:1953xh,DeWitt:1965jb,DeWitt:1967yk,DeWitt:1967ub,DeWitt:1967uc} coefficients\footnote{After Hadamard, Minakshisundaram,  and DeWitt \cite{Gibbons:1979ig,Barvinsky:2015bky}.} which are  local scalar functions of the background fields. A  general expression  in any spacetime dimension is known explicitly for the first few of them in terms of $\textbf{E}$ and geometric invariants. 

Because the first $d/2-1$ terms have negative powers of $\e$, the above short time expansion  is divergent. The divergences can be renormalized by appropriate local counterterms;  the remaining finite  piece is given by $a_{d/2}(x)$ which thus determines the Weyl anomaly through \eqref{anomaly1}.

The relevant $a_n(x)$ coefficients  up to spacetime dimension $d=4$ are given by \cite{Vassilevich:2003xt}
\bea
a_0 &=& \text{tr}\, \mathbf{1}\\
a_1 &=&  \text{tr}\left(\frac{1}{6}\,R\, \textbf{1}-\textbf{E}\right)\\
a_2 &=&\text{tr}\, \bigg( \half\,\textbf{E}^2- \frac{1}{6}\,\nabla^2 \textbf{E}-\frac{1}{6}\,R\,\textbf{E}+\frac{1}{12}\,\O_{\m\n}\,\O^{\m\n}\bigg. \nn\\
 & &\bigg.\qquad\qquad+\frac{1}{180}\,\left(6\,\nabla^2 R+\frac{5}{2}\,R^2 -\half E_4 +\frac{3}{2}\,W^{2}\right)\,\textbf{1}\bigg)\, ,
\eea
where $\text{tr}\, \textbf{1}$ traces all indices, $\nabla_\m := \partial_{\m}+ \omega_{\mu}$ is the covariant derivative involving only the spin connection, and $\O_{\m\n}=[D_\m,D_{\n\,}]$ is the field strength of the full connection. $E_{4}$ is the Euler density in four dimensions and $W^{2}$ is the square of the Weyl tensor ${W_{\m\n\r}}^\s$ defined by
\bea
E_{4} &=& R_{\m\n\r\s}R^{\m\n\r\s} -4 R_{\m\n}R^{\m\n} + R^{2} \\
W^{2} &=& R_{\m\n\r\s}R^{\m\n\r\s} -2 R_{\m\n}R^{\m\n} + \frac{1}{3} R^{2}\, .
\eea

In four dimensions, the anomaly $\cA = \cB + \cC$ is determined by $a_2(x)$.  In Table \ref{table} we list the anomalies for the operators appearing in the Yang-Mills and conformally-coupled scalar actions. 
We have dropped the terms proportional to  $\nabla^2 R$ and $ \nabla^2\varphi^2$. Such operators follow from the Weyl variation of  local terms in the action, namely $R^2$ and $R\,\varphi^{2}$, hence are not genuine anomalies  \cite{Bonora:1983ff}. The vector potential operator $\cO_A$ corresponds to the Feynman gauge $\xi=1$ and $F^{2} := F^a_{\m\n}\,F^{a\,\m\n}$. Note that the $a_{2}(x)$ coefficients for the ghost and vector operators individually contain a term proportional to $R^{2}$. This is related to the the fact that the operators are not individually Weyl covariant. However, taken together, the $R^{2}$ terms cancel from the anomaly as expected from  the Wess-Zumino consistency condition.

\begin{table}[htp]\setlength{\tabcolsep}{15pt}
\begin{center}
\begin{tabular}{|c|c|c|c|}
\hline
Field & $\cO$ & $ 16\pi^{2}  \mathcal{B}$ & $16\pi^{2} \cC $   \\ \hline
$\Phi$ & $-D^2 +\frac{1}{6}\,R$ & $-\frac{1}{12} \, F^{2}$& $\frac{N}{180}\,\left(-E_4 +3\,W^2\right)$  \\
$c, \bar c$& $-D^2$  & $\,\frac{N}{6}\, F^{2}$& $ \frac{N^2-1}{180} \,\left(-5\,R^{2}+E_4 -3\,W^{2}\right)$   \\
$A_{\mu}$ & $-D^2\, g^{\m\n}+R^{\m\n} -2\, F^{\m\n}$ & 
$\frac{5\,N}{3}\, F^{2}$& $ \frac{N^2-1}{180} \,\left(5\, R^{2}-32\, E_4 +21\,W^{2}\right)$   \\
$\Psi$& $-D^2 +\frac{1}{4}\,R - \frac{1}{2}F_{\m\n}\Gamma^{\m}\Gamma^{\n}$ & $-\frac{1}{3}\, F^{2}$&$\frac{N}{180}\,\left(-\frac{11}{2} E_4 +9\,W^2\right)$   \\
$\varphi$ & $-\nabla^2 +\frac{1}{6}\,R + \half \l \varphi^{2}$ & $\frac{1}{8}\l^{2}{ \varphi}^{4}\, $& $\frac{1}{180}\,\left(-E_4 +3\,W^2\right)$  \\
\hline
\end{tabular}
\end{center}
\caption{Weyl anomalies in $d=4$. The contributions from the complex scalars $\Phi$ and fermions $\Psi$ to the $\cB$ anomaly are different for the abelian and non-abelian cases. In the table we have indicated the non-abelian ones relevant for Yang-Mills. For quantum electrodynamics, the contributions are multiplied by a factor of two due to the choice of normalization of the non-abelian gauge group generators.}
\label{table}
\end{table}
Putting these results together, the Weyl anomaly equation for Yang-Mills is
\be\label{YMtrace}
J_{x}(\cS_{1} [g, A]) =  \frac{\delta \cS_{1} [g, A]}{\delta \Omega} =  \left(\frac{b}{4}   \,  F^{2} (x) - \cC(x)  \right)\sqrt{|g|}\, ,
\ee
with
\bea\label{qcdb}
b = \frac{1}{48\pi^2}\left(N_S + 4N_F -22N\right) 
\eea
for an $SU(N)$ theory with $N_S$ scalars and $N_F$ fermions in the fundamental. In quantum electrodynamics integrating out $N_{F}$  fermions and $N_{S}$  scalars, one would get a similar result with
\bea\label{qedb}
b = \frac{1}{24\pi^2}\left(N_S+ 4N_F\right)  \, .
\eea

For the real scalar field $\varphi$ with quartic self-interaction, we similarly obtain
\be\label{scalartrace}
J_{x}(\cS_{1} [g, \varphi]) =  \frac{\delta \cS_{1} [g, \varphi]}{\delta \Omega} = \left( -\frac{b\,\l }{4!}  \,\varphi^{4} (x)  - \cC(x) \right)\sqrt{|g|} \, 
\ee
with the beta function coefficient given by
\be\label{scalarbeta}
b = \frac{3 \l}{16\pi^{2}} \, .
\ee

\section{Nonlocal Effective Actions \label{NEA}}

In this section we derive the one-loop quantum effective actions from the anomalies following the discussion in the previous section. We drop the subscript `$1$' used earlier to indicate the one-loop results.   As a simple illustration,  we first derive  the two dimensional Polyakov action from the $\cC(x)$ anomaly.  In  four dimensions,  we  ignore the  $\cC(x)$ anomaly and focus only on the $\cB(x)$ anomaly to derive the  effective action  for the background fields $\O, A, \varphi$.

\subsection{The Polyakov action in Two Dimensions}

The trace anomaly \eqref{anomaly} for a massless  free scalar in two dimensions is given by
\be
\cA(x)=  \text{tr} \,K_{\varphi}(x,x,\e)\,.
\ee
The finite contribution to the trace in two dimensions is given by the coefficient $a_1(x)$: 
\be
\cA(x)=  \frac{1}{4\pi}\,a_1(x)= \frac{1}{4\pi}\,\text{tr}\,\left( \frac{1}{6}\,R\,\textbf{1}\right)=\frac{1}{24\pi}\,R\, .
\ee
In this case $\cB =0$ and the anomaly is purely gravitational. Using \eqref{lemmaE} and the Weyl transformation for the Ricci scalar
\be\label{Rtransform}
R = e^{-2\O}(\bar R - 2 \bar\nabla^{2}\O) \quad \textrm{for}  \quad g = e^{2\O} \bar g \, ,
\ee
 the effective action is given by $\cI[g] =\cI[\bar g] + \cI_{\cC}[\bar g, \Omega]$ with
\bea
\cI_{\cC}[\bar g, \Omega] &=& \int\limits_0^1dt\int d^2x \sqrt{|\bar g\,e^{2\,t\,\O(x)}|}\,\O(x)\, \cA[\bar g\,e^{2\,t\,\O(x)}]\non\\
&=& \frac{1}{24\pi} \int\limits_0^1dt\int d^2x \sqrt{|\bar g|}\,e^{2\,t\,\O(x)}\,\O(x)\, e^{-2\,t\,\O(x)}\left(\bar R-2\,t\,\bar\nabla^2\O(x)\right)\non\\
&=&\frac{1}{24\pi}\int d^2x \sqrt{|\bar g|}\left((\bar\nabla\O)^2+\bar R\, \O(x)\right) \, ,
\eea
which is  the Liouville action with the correct normalization. For $\bar g_{{\m\n}} = \delta_{\m\n}$, one can solve  \eqref{Rtransform} for $\Omega$ in terms of $g$ using the fact that $\bar R=0$, and
obtain  the Polyakov action
\be
\cI[g]= -\frac{1}{96\pi}\int d^2x\sqrt{|g|}\,R\,\frac{1}{\nabla^{2}}\,R\,   .
\ee
Since in two dimensions  every metric is Weyl flat, these results are valid for a general metric. 

Under a fiducial Weyl transformation \eqref{fiducial}, the Liouville action is not invariant but transforms as
\be
\cI_{\cC}[\bar g, \O]\quad  \rightarrow\quad  \cI_{\cC}[\bar g, \O] -\frac{1}{24\pi}\int d^2x \sqrt{|\bar g|}\left((\bar\nabla\zeta)^2+\bar R\,\zeta\right)\, .
\ee
However, the  $\cI[\bar g]$ also transforms as
\be
\cI[\bar g]= -\frac{1}{96\pi}\int d^2x\sqrt{|\bar g|}\,\bar R\,\frac{1}{\bar \nabla^{2}}\bar R \quad  \rightarrow\quad \cI[\bar g] + \frac{1}{24\pi}\int d^2x \sqrt{|\bar g|}\left((\bar\nabla\zeta)^2+\bar R\,\zeta\right)
\ee
ensuring the fiducial Weyl invariance of $\cI[\bar g] + \cI_{\cC}[\bar g, \Omega]$.  

\subsection{Quantum  Effective Action for Yang-Mills Theory\label{YM}}

Applying \eqref{lemmaL2} to  the $\cB$ anomaly of the Yang-Mills theory \eqref{YMtrace} in a Weyl-flat spacetime gives
\be\label{anomaly-action-YM}
\cI_{\cB}[\eta,\O, A] =-  \frac{b }{4}\int d^{4}x \,  \,  \eta^{\r\a}\, \eta^{\s\b} \, \,  F^{a}_{\r\s}(x)  \, \Omega (x) \, F^{a}_{\a\b}(x)\, .
\ee
The flat space action can be easily determined from standard computations and is given by
\be\label{flat-action-YM}
\cI[\eta, A] =  - \frac{1 }{4\,e^{2}(M)}\int d^{4}x \,d^{4}y  \,  \,  \eta^{\r\a} \eta^{\s\b} \, \, \,   F^{a}_{\r\s}(x)  \, \langle x |\left[1 -\frac{b}{2}\,e^2(M)\, \log \left(\frac{-\partial^{2}}{M^{2}}\right) \right] |y \rangle \, F^{a}_{\a\b}(y)
\ee 
where $-\partial^{2}$ is the flat-space d'Alembertian. The kets  $|x\rangle$ here are normalized as in \eqref{Tr} but now with the flat metric $\eta$. 
The logarithm of  an operator is defined by the spectral representation
\be\label{log-operator}
\log \left(\frac{\cO}{M^{2}}\right) = \int_{0}^{\infty}d\m^{2}\, \left( \frac{1}{M^{2} + \m^{2}}- \frac{1}{\cO + \m^{2}} \right) \, .
\ee
For the flat space d'Alembertian the logarithm can also be defined by a  Fourier transform:
\be
\langle x |\log \left(\frac{-\partial^{2}}{M^{2}}\right) |y\rangle= \int \frac{d^{4 }p}{(2\pi)^{4}}e^{-i p \cdot (x -y)}\, \log \left(\frac{p^{2}}{M^{2}}\right)\, .
\ee
Putting the two things together in \eqref{lemmaL} we conclude
\be\label{curved-action-YM}
\cI[g, A] =  - \frac{1 }{4e^{2}(M)}\int d^{4}x \,  \,  \eta^{\r\a} \eta^{\s\b} \, \, \,   F^{a}_{\r\s}(x)  \, \left[1 - \frac{b}{2}\,e^2(M)\, \log \left(\frac{-\partial^{2}}{M^{2}}\right)  +b\, e^2(M) \,\Omega(x) \right]\, F^{a}_{\a\b}(x)
\ee 
where the logarithmic operator is to be understood as a bilocal expression  integrated over $y$ as in \eqref{flat-action-YM}. 
There is a  gravitational piece coming from the $\cC$ anomaly which we do not discuss. 

Note that the action  \eqref{anomaly-action-YM} arising from the anomaly follows from the local Schwinger-DeWitt expansion and does not require any weak-field approximation. Thus, the main limitation in computing \eqref{curved-action-YM} comes from the evaluation of the flat space action \eqref{flat-action-YM}.
In \eqref{flat-action-YM} we have used the weak gauge field approximation $F^{4} \ll \nabla^{4} F^{2}$ as one normally does in flat space quantum field theory.
It may be possible to compute the flat space action in  other regimes, for example, in the regime of constant field strength. This can extend the range of validity of our results.

It is instructive to deduce this result  using dimensional regularization. 
Again, the classical action (or the bare action in the UV) is  given by
\be\label{YMclassical}
\cI_{0}[g, A]= -\frac{1}{4 e_{0}^{2}}\,  \int d^{4}x \sqrt{|g|}  \, g^{\r\a} g^{\s\b} \, {  F^{a}_{\r\s}  \, F^{a}_{\a\b} \,}\, .
\ee 
The classical energy momentum tensor  
\bea\label{em-tensor}
T_{\mu\nu}^{cl}= \frac{1}{e_0^2}\left(F^{a}_{\mu\sigma}F^{a\ \sigma}_{\nu}-\frac{1}{4}g_{\mu\nu}F^2\right)  
\eea
is traceless.
At the quantum level, the nonzero beta function implies   a quantum violation of Weyl invariance. For a manifestly gauge-invariant  computation of this Weyl anomaly we use dimensional regularization.  In $4-\varepsilon$ dimensions, the bare coupling $e_0$ is related to the  coupling $e$ renormalized at scale $M$ by
\bea
\frac{1}{ e_{0}^{2}} = M^{-\varepsilon}\left(\frac{1}{ e^{2}} - \frac{b}{\varepsilon}\right) \, ,  \qquad M \frac{d e^{-2}}{dM} = - b \, .
\eea
There is the usual   pole  coming from loop integrations of  quantum fluctuations around a background field. For a Weyl-flat metric,   the dimensionally regularized background field action depends only on the background gauge field and the Weyl factor $\O$:
\be
\cI^{\varepsilon}[\O, A] = -\frac{1}{4}\int d^{4-\varepsilon}x \sqrt{|\eta|}  \, \eta^{\r\a} \eta^{\s\b} \, e ^{-\varepsilon \O}\, M^{-\varepsilon} ( \frac{1}{ e^{2}} - \frac{b}{\varepsilon} ) \,    F^{a}_{\r\s}  \, F^{a}_{\a\b} \, .
\ee 
This implies that the Weyl variation of the renormalized effective action for the background field is no longer zero and is given by
\be
\d \cI[\eta,\O, A] = -\frac{b}{4}  \int d^{4}x \sqrt{|\eta|}  \, F^{2}  \, \d \Omega \, ,
\ee
consistent with the results obtained using the proper time regularization. 

\subsection{Quantum  Effective Action for a Self-interacting Scalar Field\label{Scalar}}

For a conformally coupled scalar field $\varphi$, one can similarly determine the one-loop effective action $\cI[g, \varphi]$.  Integrating the $\cB$ anomaly of \eqref{scalartrace} gives
\be\label{scalarB}
\cI_{\cB}[\eta, \Omega, \bar \varphi] =  \frac{b\,\l}{4!} \int d^{4}x \,  \bar\varphi^{4}(x) \, \O (x) \, , 
\ee
with $b$ given by \eqref{scalarbeta} and $\lambda$ being the renormalized quartic coupling defined at the scale $M$. 
The flat space action obtained from standard computations gives
\be\label{q-action-scalar}
\cI[\eta, \bar\varphi] =  - \int d^{4}x  \left[  \frac{1 }{2} |\partial \bar\varphi |^{2}+ \frac{\l}{4!} \bar\varphi^{2}(x)  \left(1+ \frac{b}{2}\log \left(\frac{-\partial^{2}}{M^{2}}\right) \right)\bar\varphi^{2}(x) \right] \, .
\ee 
Using \eqref{lemmaL}, the full effective action is given by
 \be\label{q-action-scalar2}
\cI[g, \varphi] =  - \int d^{4}x  \left[  \frac{1 }{2} |\partial \bar\varphi |^{2}+ \frac{\l}{4!} \bar\varphi^{2}(x)  \left(1+ \frac{b}{2}\log \left(\frac{-\partial^{2}}{M^{2}}\right) - b\,\O(x) \right)\bar\varphi^{2}(x) \right] \, .
\ee 

As in the case of the Yang-Mills action,  the part of the action  \eqref{scalarB} arising from the anomaly  does not require any weak-field approximation and is exact. The flat space action \eqref{q-action-scalar} is valid only assuming rapidly varying field. It could be evaluated though in other regimes  of interest using techniques such as the large proper time expansion developed in \cite{Barvinsky:2002uf,Barvinsky:2003rx} or the Coleman-Weinberg method. However, note 
that when the field $\varphi$ is in the Coleman-Weinberg regime,  the field $\bar{\varphi}$ may not be unless the scale factor is also slowly varying.
 
We see that the net effect of the Weyl anomaly in the combined action is to change the renormalization scale to an effective \textit{local} renormalization scale $M(x) := Me^{\Omega(x)}$ consistent with \eqref{Mrelation}. One can explain the answer intuitively if the scale factor is varying slowly  compared to the typical scale of field variations (for example in a particle physics experiment in an expanding universe). In this case, one can use local momentum expansion to write $-\partial^{2} = k^{2}$.  In local experiments  \eqref{q-action-scalar2} can be interpreted as a flat space action with momentum-squared $k^{2}$ but with a  position dependent cutoff $M(x)$. 
One can equivalently interpret $k^{2}/M^{2}(x)$ as $p^{2}(x)/M^{2}$ in terms of physical momentum-squared  $p^{2}= e^{-2\Omega(x)}k^{2}$ with a fixed RG scale $M$.
This suggests that we can use the local renormalization group to define a position-dependent `running' coupling
\bea\label{running}
\l(p^{2}(x)) &=& \frac{\l}{1- \frac{b}{2}\log \left(\frac{p^{2}(x)}{M^{2}}\right)} \, \\
&=& \l \left[ {1+  \frac{b}{2}\log \left(\frac{p^{2}(x)}{M^{2}}\right)}+ \frac{b^{2}}{4}
\log ^{2}\left(\frac{p^{2}(x)}{M^{2}}\right) + \ldots\right] \, .
\eea
Equation \eqref{running} re-sums the leading logarithms to all orders as with the usual renormalization group but now locally.
The  effective coupling  decreases as the universe expands because the beta function is positive. Consequently, the renormalization-group improved answer becomes better and better at late times even though naive perturbation theory would break down. 
The local renormalization group thus extends the range of applicability of the perturbative computations. 

In more general situations with a rapidly varying scale factor, one cannot use the momentum basis as above but equation \eqref{q-action-scalar2} is still valid. One might be tempted to interpret the full answer in terms of the logarithm of the covariant d’Alembertian in curved spacetime, $\log \left({-\nabla^{2}}/{M^{2}}\right)$. However, the full covariantization is rather nontrivial and requires many more nonlocal covariant terms which combine into a Weyl-invariant piece \cite{Barvinsky:1995it, Donoghue:2015nba}.  We discuss this in detail in $\S\ref{BV}$. 

\subsection{Equations of Motion \label{EM}}

If we are interested in the equations of motion of the fields in a fixed background  metric, then the metric does not need to be varied and can be assumed to be Weyl flat. 
The equations of motion for the background Yang-Mills field follow straightforwardly from the action $\cI[g, A]$ \eqref{curved-action-YM} and are given by
\be
\frac{\d\cI[\eta, A]}{\d A_{\m}} + \frac{\d\cI_{\cB}[\eta, \Omega, A]}{\d A_{\m}} = 0\, .
\ee
The first term gives the logarithmic modifications to the flat space equations of motion arising from integrating out massless charged particles. The second term gives rise to the anomalous coupling to the conformal factor of the metric which breaks the Weyl invariance. Similar considerations extend to the equations of motion for the Weyl-transformed scalar $\bar\varphi$.

These  actions are thus adequate for studying the  equations of motion for the  fluctuations of the gauge field or a scalar field in an arbitrary Robertson-Walker background including the full anomalous dependence on the Weyl factor. This is the situation one encounters, for example, in studying the primordial perturbations of a scalar or of the electromagnetic field in a slowly rolling inflationary background.

\subsection{Barvinsky-Vilkovisky Expansion and Conformal Decomposition\label{BV}}

In the weak curvature limit, we can compare our results with the covariant curvature expansion developed by Barvinsky, Vilkovisky, and collaborators \cite{Barvinsky:1984jd,Barvinsky:1985an, 
Barvinsky:1988ds,Barvinsky:1994hw,Barvinsky:1994cg,Barvinsky:1995it}.
It  provides a useful check on our results obtained using a rather different method which does not rely on the weak curvature approximation. 

The main idea behind the Barvinsky-Vilkovisky (BV) expansion is to  decompose the metric as $g_{\m\n}= \eta_{\m\n} + h_{\m\n}$ and  treat the fluctuations $h_{\m\n}$    as perturbations. The heat equation satisfied by the kernel $K(s)$  can  be solved perturbatively around flat space using the analog of  the Dirac interaction picture in quantum mechanics. The perturbative answers are then `covariantized’ up to a given order to express them in terms of covariant derivatives  and polynomials of generalized curvature tensors  schematically denoted as $\mathcal{R}$, which includes both terms like  $R_{\mu\nu}$ as well as $F_{\mu\nu}$.  This expansion is  valid for small generalized curvatures but for the entire range of the  proper time $s$. The effective action can thus  be obtained by evaluating the  integral \eqref{trlog2}. The final answer can be expressed in terms of non-local `form factors’ and schematically takes  the  form
\be\label{BVresult}
\cS =  \cS_0+ \cS_1^{loc}+\int d^dx\sqrt{|g|}\  \sum_{i=1}^{5}f_i(-\nabla^2_2)\mathcal{R}_1\mathcal{R}_2 + \sum_{i=1}^{29}\mathcal{F}_i(-\nabla^{2}_1,-\nabla^{2}_2,-\nabla^{2}_3)\mathcal{R}_1\mathcal{R}_2\mathcal{R}_3(i) +\mathcal{O}(\mathcal{R}^4) \, .
\ee
The notation is a shorthand for terms containing all possible combinations of curvatures such as $Rf_{RR}(-\nabla^2)R$ and $\mathcal{F}_2(-\nabla^2_1,-\nabla^2_2,-\nabla^2_3)F^{\ \mu}_{1\ \nu}F^{\ \nu}_{2\ \sigma}R^{\ \sigma}_{3\ \mu}$, for example.
The form factors $f$ and $\mathcal{F}$ as functions of the covariant Laplacian are generically non-local operators,  and are to be understood as properly convoluted with the functions they act upon \cite{Barvinsky:1985an}.
A similar result has been obtained for massless quantum electrodynamics by a somewhat different method   by Donoghue and El-Menoufi \cite{Donoghue:2015nba,Donoghue:2015xla} by evaluating the one-loop Feynman diagrams for small metric fluctuations around flat space and then covariantizing the answers. 

An important advantage of this expansion is that it gives all nonlocal terms in the action directly to a given order in perturbation theory. The  price to pay though is that these expressions are necessarily perturbative, valid only in the regime of 
$\cR^{2} \ll \nabla^{2} \cR  $. 
Note that there are two perturbative expansions at work. The loop expansion parameter is $e^2$ or $\lambda$, while the BV expansion involves a further  approximation which treats the field perturbations, such as $h_{\m\n}$, $A_\mu$ or $V''$ as small. This \textit{weak field} approximation implies that  terms of the form $\partial^{2} h \partial^{2} h$ are  to be regarded as much smaller than terms of the form $\partial^{4}h$ even though both have the same number of derivatives. Upon covariantization, it implies that the BV expansion is valid if $ R^2 \ll \nabla^2 R$. By contrast, the local Schwinger-DeWitt expansion is valid for short proper time $\e R \ll 1$ or equivalently for the  entire weak gravity regime $R \ll M_{0}^{2}$  without any further restrictions on curvatures. 


To compare \eqref{BVresult} with our results, it is necessary to go to  third order in the BV expansion.  Explicit expressions to this order have been worked out  in  \cite{Barvinsky:1994hw} but they are rather  complicated going over several pages. It is not immediately  obvious how these expressions  could reduce to the simple expressions that we obtained earlier. However,  one can use the fact that the Weyl variation of the  BV effective action must correctly reproduce the \textit{local} Weyl anomaly. 
This observation suggests a `conformal decomposition’ of the action in terms of a Weyl-invariant piece and a Weyl-variant piece \cite{Barvinsky:1994cg,Barvinsky:1995it,Donoghue:2015nba}. This conformal decomposition is what is most easily compared with our results. 

To illustrate the idea, consider the BV effective action for quantum electrodynamics   obtained by integrating out massless charged fields in the presence  a background gauge field $A$.  To third order in curvatures it is given by \cite{Donoghue:2015nba}:
\bea\label{qed-result}
\cI[g, A] &=& -\frac{1}{4}\int d^4x \sqrt{|g|} \left\{ \frac{1}{e^2}F_{\mu\nu}F^{\mu\nu} - \frac{b}{2} \, \left[F^{\mu\nu}\log\left(\frac{-\nabla^2}{M^2}\right)F_{\mu\nu}\,+\frac{1}{3}\, F^2\frac{1}{\nabla^2}R\,\,+\right. \right.  \nonumber \\ 
& +& 4R^{\mu\nu}\frac{1}{\nabla^2}\left(\log\left(\frac{-\nabla^2}{M^2}\right)\left(F^{a}_{\mu\sigma}F^{a\ \sigma}_{\nu}-\frac{1}{4}g_{\mu\nu}F^2\right) - F_{\mu\sigma}\log\left(\frac{-\nabla^2}{M^2}\right)F_{\nu}^{\sigma} \,\, + \right. \nonumber \\ 
&+& \left.\left.\frac{1}{4}g_{\mu\nu}F^{\alpha\beta}\log\left(\frac{-\nabla^2}{M^2}\right)F_{\alpha\beta}\right)-\frac{1}{3}RF^{\mu\nu}\frac{1}{\nabla^2}F_{\mu\nu}+W^{\alpha}_{\ \beta\mu\nu}F_{\alpha}^{\ \beta}\frac{1}{\nabla^2}F^{\mu\nu}\right]\, +\nonumber \nonumber \\ 
&+& \left. 4 \,\tilde b  \, F^{\mu\nu}F_{\alpha}^{\ \beta}\frac{1}{\nabla^2}W^{\alpha}_{\ \beta\mu\nu} \right\} +\mathcal{O}(\mathcal{R}^4)
\eea
where the logarithm of the covariant d’Alembertian $\log (-\nabla^{2}/M^{2})$ is defined as in \eqref{log-operator}
and 
\bea
b = \frac{1}{24\pi^2}\left(N_S+ 4N_F\right) \, , \qquad \tilde b  = \frac{1}{96\pi^2}(-N_{S} + 2 N_{F})\, .
\eea
Note that  $b$ is the usual beta function coefficient \eqref{qedb} in flat space but $\tilde b$ is relevant only in curved backgrounds.
We have ignored the purely gravitational terms coming from the $\cC$ anomaly that are  independent of the background gauge field. 

It turns out that except for the second term in the square bracket,  all other terms in \eqref{qed-result} are actually Weyl invariant \cite{Barvinsky:1995it, Donoghue:2015nba}.  This `conformal decomposition'  then  implies that the only Weyl-variant term that could contribute to the $\cB$ anomaly is  precisely this second term:
\be\label{BV-anomaly-action} 
\tilde\cI_{\cB}[g, A]=-\frac{b}{4}\int d^4x\sqrt{|g|}\ F^{\mu\nu}\left(-\frac{1}{6}\frac{1}{\nabla^{2}}R\right)F_{\mu\nu}  \, .
\ee
Since all other terms taken together are Weyl invariant, for a Weyl-flat metric they must reduce to the one-loop effective action on  flat space  \eqref{flat-action-YM}:
\be \cI[{\eta,A}] = -\frac{1}{4} \int d^4x\,\, F^{\mu\nu}\left[\frac{1}{e^2(M)}-\frac{b}{2} \log\left(\frac{-\partial^2}{M^2}\right)\right]F_{\mu\nu} \, .
\ee
Hence for a Weyl-flat metric the  action \eqref{qed-result} simplifies dramatically to 
\be \label{qedfinal}
\cI[g, A] = \cI[\eta,A] + \tilde \cI_{\cB}[\eta, \O, A] \, . \ee
We would like to compare this result with the one obtained by integrating the anomaly:
\be \label{ouraction}
\cI[g, A] = \cI[\eta,A] +  \cI_{\cB}[\eta, \O, A] \, , \quad \text{with}\quad 
\cI_{\cB}[\eta, \O, A]  =-\frac{b}{4} \int d^4x\,\eta^{\m\r}\,\eta^{\n\s}\, F_{\m\n}\,\O(x) \, F_{\r\s} \, .
\ee

To this end, we note  that the  Weyl factor $\Omega[g](x)$ can be expressed as a nonlocal covariant functional of the metric \cite{Fradkin:1978yf, Paneitz:2008, Riegert:1984kt} given by
\begin{equation}\label{Omega}
 \O[g](x)= \frac{1}{4}\int\! d^4 y\,\sqrt{|g|}\, G_4(x,y)\, F_4[g](y) \, ,
\end{equation}
where 
\begin{align}\label{F4}
F_4[g] := E_4[g]-\frac{2}{3} \nabla^2 R[g]\,  = (R_{\m\n\r\s} R^{\m\n\r\s} - 4 R_{\m\n}R^{\m\n} +        R^{2}-\frac{2}{3} \nabla^{2} R )[g] \, ,
\end{align}
and the Green function $G_{4}(x,y)$ defined by 
\be
\Delta^{x}_4[g] G_{4}(x, y) = \d^{(4)}(x, y) :=\frac{ \d^{(4)}(x -y) }{\sqrt{|g|}}\, 
\ee
is the inverse of the Weyl-covariant quartic differential operator 
\begin{equation}\label{diff-op}
 \Delta_4[g]= \left( \nabla^2 \right)^2 + 2 R^{\m\n}  \nabla_{\m}\nabla_{\n} + \frac{1}{3}\left(\nabla^{\n}R\right)\nabla_{\n} - \frac{2}{3}R\, \nabla^2 \, . 
\end{equation}
The expression \eqref{Omega} follows from the fact that for metrics  related by a Weyl rescaling $g_{\m\n}=e^{2\O(x)}\, {\eta}_{\m\n}$,
the corresponding $F_{4}$ scalars are related by
\begin{equation}\label{ftrans}
 F_4[ g] = e^{-4\O} \left( F_4[ \eta] +4  \, \Delta_4 [ \eta]\, \O \right)\, ,
\end{equation}
and the operators $\Delta_{4}$ by
\begin{equation}
\Delta_4[g]= e^{-4\O}\, \Delta_4[ \eta] \, .
\end{equation}
Since the Minkowski  reference metric satisfies
$F_4[ \eta]=0$ the  expression \eqref{Omega} follows from  inverting \eqref{ftrans}. 
This expression is manifestly covariant but nonlocal, consistent with the fact that  the anomalous $\O$ dependence  represents genuine long-distance quantum effects that cannot be removed by counter-terms that are local functionals of the metric.

When $ R^2 \ll \nabla^2 R $ one can expand the expression for $\O$ \eqref{Omega}  in curvatures to obtain
\be {\O}[g] (x)= -\frac{1}{6}\frac{1}{\nabla^{2}}R + \mathcal{O}(R^2) \ee
which when substituted in \eqref{ouraction} reproduces the anomaly action obtained in the BV regime \eqref{BV-anomaly-action}.

To recover the full expression for $\O$ in the Barvinsky-Vilkovisky formalism one must invert the
operator $\Delta_{4}$ perturbatively, which involves higher and higher orders in curvatures. As a result,  the expression  \eqref{Omega} for $\Omega$ will similarly involve terms to arbitrary order in the curvature expansion. 
This implies that to recover the exact and simple expression \eqref{curved-action-YM} obtained by integrating the Weyl anomaly it would be necessary to re-sum the covariant perturbation theory  \eqref{qed-result} \textit{to all orders} in curvatures $R$ for the class of Weyl-flat metrics. Since $\cI_{\cB}$  already contains  $F^{2}$, the next correction  is of order $F^{2}R^{2} \sim \cR^{4}$ in the generalized curvature expansion.
In other words, 
\be
\cI_{\cB}[\eta, \O, A] = \tilde \cI_{\cB}[\eta, \O, A]  + \cO (\cR^{4})\, .
\ee 
Already at order $\cR^{4}$, the expression in the BV expansion becomes unmanageable. It is remarkable that the simple expression \eqref{lemmaL} re-sums this expansion to all orders albeit for a restricted class of Weyl-flat metrics. 

Thus,  explicit `covariantization' of our answer obtained by integrating the anomaly can lead to rather complicated expressions even though the exact answer \eqref{ouraction}
is strikingly simple. As noted earlier, our procedure guarantees that the full answer depends only the physical metric $g$ even though \textit{a priori} the right hand side appears to depend  on $\eta_{\m\n}$ and $\O$ separately.

\subsection*{Acknowledgments}

A major part of this work was conducted within the framework of  the ILP LABEX (ANR-10-LABX-63)  supported by French state funds managed by the Agence National de la Recherche within the Investissements d'Avenir programme under reference ANR-11-IDEX-0004-02,  and by the project QHNS in the program ANR Blanc SIMI5. T.B. and A.D. acknowledge the support from the Indo-French Centre CEFIPRA, under project number 5204-4. T. B. especially thanks the ICTP and the three host Indian institutes under the CEFIPRA project (the Department of Theoretical Physics at the Tata Institute of Fundamental Research, the International Centre for Theoretical Sciences, and the Harish-Chandra Research Institute) for the hospitality and stimulating environment provided during part of this work.  We thank  Alba Grassi, Takeshi Kobayashi, Roberto Percacci, Adam Schwimmer, and Ashoke Sen for useful discussions.

\bibliographystyle{JHEP}
\bibliography{weyl}
\end{document}